\documentclass[aps,pra,twocolumn]{revtex4-2}
\usepackage{amsmath,epsfig,mathrsfs,graphicx,amssymb,color}
\usepackage[T1]{fontenc}
\usepackage{t1enc}
\usepackage{hyperref}
\usepackage{array}
\usepackage{booktabs}

\def\be#1\ee{\begin{equation}#1\end{equation}}
\newcommand{\ba}{\begin{eqnarray} }
\newcommand{\ea}{\end{eqnarray} }

\usepackage[cp1250]{inputenc}   % pozwala wprowadzać polskie znaki bezpośrednio z klawiatury
\usepackage{amsmath}            % AMS-LaTeX (AMS-Math)
\usepackage{amsfonts}           % ekstra fonty matematyczne
\usepackage{amssymb}            % ekstra symbole matematyczne
\usepackage{enumerate,graphicx}
\usepackage{array}
\usepackage{booktabs}

\usepackage{physics}

\usepackage{amsmath}            % AMS-LaTeX (AMS-Math)
\usepackage{amsfonts}           % ekstra fonty matematyczne
\usepackage{amssymb}            % ekstra symbole matematyczne
\usepackage{enumerate,graphicx}
\usepackage{array}
\usepackage{booktabs}
\usepackage{yquant}
\usetikzlibrary{fit, quotes}
\usetikzlibrary{snakes,arrows,shapes}
\useyquantlanguage{groups}
\usepackage{physics}

\def\mb{\begin{pmatrix}}
\def\me{\end{pmatrix}}
\def\be#1\ee{\begin{equation}#1\end{equation}}

\def\mb{\begin{pmatrix}}
\def\me{\end{pmatrix}}
\def\be#1\ee{\begin{equation}#1\end{equation}}

\begin{document}

\title{Quantum null-hypothesis device-independent Schmidt number witness}

\author{Josep Batle$^{1,2}$}
\email{jbv276@uib.es, batlequantum@gmail.com}         
\author{Tomasz Bia{\l}ecki$^{3}$}
\author{Tomasz Rybotycki$^{4,5,6}$}
\author{Jakub Tworzyd{\l}o$^{3}$}
\author{Adam Bednorz$^{3}$}
\email{Adam.Bednorz@fuw.edu.pl}

\affiliation{$^1$
Departament de Fisica and Institut d'Aplicacions Computacionals de Codi Comunitari (IAC3),
Campus UIB, E-07122 Palma de Mallorca, Balearic Islands, Spain}

\affiliation{$^2$CRISP -- Centre de Recerca Independent de sa Pobla, 07420, sa Pobla, Balearic Islands, Spain}
\affiliation{$^3$Faculty of Physics, University of Warsaw, ul. Pasteura 5, PL02-093 Warsaw, Poland}
\affiliation{$^4$Systems Research Institute, Polish Academy of Sciences, 6 Newelska Street, PL01-447 Warsaw, Poland}
\affiliation{
	$^5$Nicolaus Copernicus Astronomical Center, Polish Academy of Sciences, 18 Bartycka
	Street, PL00-716 Warsaw, Poland
}
\affiliation{
	$^6$Center of Excellence in Artificial Intelligence, AGH University,
	30 Mickiewicza Lane, PL30-059 Cracow, Poland
}

\begin{abstract}
We investigate the dimensionality of bipartite quantum systems by construction of a device-independent null witness test. This test assesses whether a given bipartite state conforms with the expected quantum dimension, Schmidt number, and distinguishes between real and complex spaces. By employing local measurements on each party, the proposed method aims to determine the minimal rank. By performing an experimental demonstration on IBM Quantum devices, we prove the exceptional accuracy of the test and its usefulness in diagnostics beyond routine calibrations.
 One of the tests shows agreement with theoretical expectations within statistical errors. However, the second test failed by more than 6 standard deviations, indicating unspecified parasitic entanglements, with no known simple origin. 

\end{abstract}

\maketitle

\section{Introduction}
Quantum bipartite systems differ from their classical counterparts, especially when entangled.
They violate local realism \cite{epr,bell,chsh,chi,eber} and are useful in quantum computation, steering and teleportation \cite{steer,steer2,tele,tele2}.
The most common Bell entangled state involves both two-level parties. 
In this case, correlations can be predicted from the knowledge of the state and available measurements.

On the other hand, it is important to assess the dimension of the system in question (quantum or classical)
for the error correction and mitigation tools to work, assuming restricted Hilbert spaces \cite{mit1,mit2,mit3}. So far, such tests involved a single
party, when a protocol of independent sets of preparations and measurements allowed construction of a dimension witness, initially in the form of inequalities
\cite{gallego,dim1}, also tested experimentally \cite{hendr,ahr,ahr2,leak}.
Analogous tests, as inequalities, have been proposed for bipartite states \cite{bell-dim,bell-dim2,bell-dim3,bell-dim4,bell-dim5}
based on families of Bell-type inequalities \cite{cglmp}.
If the contribution from extra states is small it is better to seek a null test (a function of probabilities is exactly equal to zero), for instance the Sorkin identity \cite{sorkin} in the three-slit experiment \cite{tslit,btest1,btest2}  verifying Born's rule \cite{born} under certain assumptions.
The quantum dimension can also be tested by null hypotheses \cite{dim,chen,bb22}, demonstrated experimentally \cite{opt,ibm}.
Testing an exact value, up to statistical error, boosts the accuracy of the test.

In this work, we propose the test if a bipartite state is of the expected quantum dimension as a null hypothesis, based on independent measurements of each party,
i.e. a measurement-measurement scenario with a single, common preparation, in contrast to the previous
 preparation-measurement protocol \cite{dim,opt}.
We construct a  witness, function of bipartite probabilities, which is zero if any party can be represented in the space of the expected dimension, depending on whether the space is complex or real
\cite{real}. The measurements can be arbitrary, device-independent, but must be performed in local subspaces. In other words, we test the \emph{minimal} bipartite space dimension to represent the state, called Schmidt number \cite{sch1,sch2}.
Existing witnesses of the Schmidt number \cite{wit1,wit2,wit3,wit4,brun,vert,vert2,vert3,vert5,vert6,wit5} are based on inequalities, often only state-independent,
or only narrowly violated in a larger space.
Note that, like Bell-type test, any violation requires nonclassical states for linear inequalities, and quite faithful implementation of quantum operations on physical devices.
Therefore, linear inequalities, although robust against calibration changes, are in principle less accurate and less general than null witnesses,
as Schmidt number is independent of nonclassicality.
Then the larger dimension is usually already clear by the implementation itself. On the other hand, a null device-independent witness is useful
when the space is trusted to be restricted and one has no access to accurate operations, which can be noisy. It is a precise tool to certify the device solely with respect to its Hilbert dimension, not errors within the space.

Our witness tests if an entangled state has the expected Schmidt number $d$ by (a) $n$ measurements  chosen by each of the two parties or (b)
single measurements with $n+1$ outcomes. The measured observables can be arbitrary and their representation is irrelevant as long as they are local.
The witness is the determinant of the probability matrix \cite{bell-det}, of size depending on the dimension, becoming $0$ for sufficiently large $n$,
up to a finite statistics error. 
Moreover, we demonstrate the feasibility of the test on qubits on IBM Quantum, if a created entangled state fits in the $2\times 2$ qubit-qubit bipartite space. 
The results of the test (a) are in agreement with the null hypothesis within the corresponding statistical errors. 
The locality of measurements implies that the parties cannot affect each other, so one can perform simultaneously the sanity check of no-signaling, common also in experimental Bell-type tests \cite{hensen,vien,nist,munch}. However, the test (b) failed by more than 6 standard deviations. Taking into account 
the robustness of our
test against local errors and negligible leakage and crosstalk contribution, it shows extreme accuracy of our test, revealing problems beyond standard diagnostics,
which demand urgent explanation, either technical or fundamental.

\begin{figure}
\includegraphics[scale=.3]{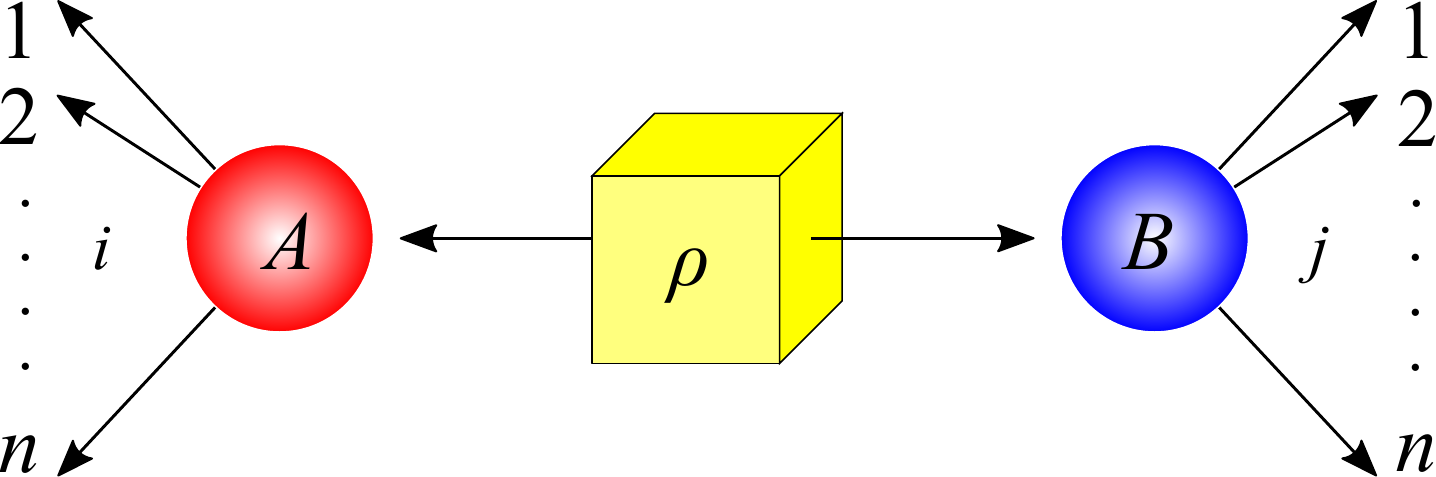}
\caption{Two parties $A$ and $B$ perform either (a) one of $n$ independent measurements indexed by $i$ and $j$ or (b) single measurements with $n$ outcomes, also indexed by $i$ and $j$, on the shared bipartite states $\rho$. The case (b) differs from (a) by imposed condition that outcomes sum up to identity measurement. }
\label{sab}
\end{figure}

\section{The witness}
Suppose we have a composite (tensor) system of $A$ and $B$ and the initially prepared state $\rho=\mathcal E_A\mathcal E_B\tilde{\rho}$, where $\mathcal E$ are local maps in respective subspaces \cite{nielsen}. To shorten notation, we shall drop the tensor sign whenever unambiguous, i.e. $\mathcal E_A\mathcal E_A\equiv \mathcal E_A\otimes \mathcal E_B$ and $AB\equiv A\otimes B$.
Assuming the form
\be
\rho=\sum_{k}|\psi_k\rangle\langle\psi_k|\label{decomp}
\ee
with states $|\psi_k\rangle$,
we can purify $\tilde{\rho}$ by adding the auxiliary system $C$, so that $\tilde{\rho}=|\psi\rangle\langle\psi|$ with 
$|\psi\rangle=\sum_k|\psi_k,k\rangle$, so that $k$ labels the basis states in $C$. 
The Schmidt number \cite{sch1,sch2} is the minimal possible number $d$ of basis states in the Schmidt decomposition in the respective spaces $A$ and $BC$, 
\be
|\psi\rangle=\sum_{k'=1}^d|k'_Ak'_{BC}\rangle.
\ee
In the above purification $\mathcal E_A$ is the identity, while $\mathcal E_B=\mathrm{Tr}_C$ (tracing out $C$).

We define the probability 
\be
p_{ij}=\mathrm{Tr}A_iB_j\rho=\langle\psi|\tilde{A}_i\tilde{B}_j|\psi\rangle,
\label{pij}
\ee
 with $\tilde{A}_i=\mathcal E^\dag_A A_i$, $\tilde{B}_j=\mathcal E_B^\dag B_j$ for the observables $A_i$, $B_j$ and local maps $\mathcal E_A$, $\mathcal E_B$
 for the above constructed Schmidt decomposition states $|\psi\rangle$. Replacing $\tilde{A}_i\to A_i$, $\tilde{B}_j\to B_j$,
and $|\psi\rangle\langle\psi|\to\rho$, we can reduce the whole discussion of the Schmidt number to the $d\times d$ dimensional composite Hilbert space.
Let us construct lists of local Hermitian observables (a) $0\leq A_i,B_j\leq 1$ for $i,j=1\dots n$ with yes/no or 1/0 outcome for $n$ independent measurements
and auxiliary $A_0=B_{0}=1$,
or (b) $0\leq A_i,B_j$ for $i,j=1\dots n+1$ and $\sum_i A_i=\sum_i B_i=1$ for single measurements with $n+1$ outcomes,
 see Fig. \ref{sab}. 
 In the case (a) the actual number of measurements is $n\times n$.
Then the first row and column ($0$) do not need a separate measurement, as one can simply discard the outcome of the other party from
measurements already done. Nevertheless, these entries
must satisfy no-signaling (independence of the measurement on the other party),
 i.e. $p_{i0}$ (or $p_{0j}$) is obtained in the measurement $(i,j)$  but cannot depend on $j$  (or $i$). The corner element is constant $p_{00}=1$.
In both cases, our witness is the $(n+1) \times (n+1)$ determinant $W_n=\det p$ which is equal $0$ if the Schmidt number satisfies $d^2\leq n$ (complex) or $d(d+1)/2\leq n$ (real)
because the size of $p$ exceeds the maximal rank of the set of allowed matrices $A_i$ or $B_j$, which span the available linear space.
If the dimension of the linear space of observables is smaller than the size of the matrix then some observable must be a linear combination of the rest.
By linearity of the matrix (\ref{pij}) as a function of observables, the same applies to its corresponding column or row, and the determinant must vanish.
The real symmetric matrix $d\times n$ is represented by $d(d+1)/2$ independent real numbers. A complex Hermitian matrix has additionally $d(d-1)/2$
real independent numbers representing imaginary antisymmetric matrices. The test is device-independent, we make no assumption about the actual realization of measurements, that is the test does not rely on the
 mathematical model of the observables.

This happens in the case $d=2$ real for $n\geq 3$, and $d=2$ complex for $n\geq 4$ which is the case we test.
For instance, if $W_4\neq 0$ then  either we have (i) a quantum system of $d=3$ or (ii) a classical system of $d=5$.
This is why our test is not intended to check just whether the system is classical or quantum but rather the value of $d$, depending on the type of the system.

\begin{table}
\begin{tabular}{c|*{7}{c}}
\toprule
$n\backslash d$&$2$&$3$&$4$&$5$&$6$&$7$&$8$\\
\midrule
$1$&$1$&&&&&&\\
$2$&$0$&$0.59$&$1$&&&&\\
$3$&$0$&$0$&$1$&&&&\\
$4$&$0$&$0$&$0$&$0.74$&$0.76$&$0.79$&$1$\\
$5$&$0$&$0$&$0$&$0$&$0.55$&$0.59$&$1$\\
\bottomrule
\end{tabular}
\caption{The calculated maxima of the witness quantity $4^nW_n$ in the classical approach for $d=2\dots 8$ and number of measurements 
$n=1\dots 5$. Entries in the empty cells saturate the bound at 1.}
\label{tbc}
\end{table}

Even if the Schmidt number is larger than expected, the witness can remain zero by accident. 
Nevertheless, the absolute upper bound on $W_n$, the same in the classical and quantum case
(see detailed proof in Appendix \ref{appa}), allows us to estimate how large the witness can be if the expected $d$ is exceeded.
To determine it over possible states and measurements, let us consider a simpler, classical case.
Suppose the composite system has the states $(i',j')$ for the parties $A$ and $B$, $i',j'=1\dots d$. Then
the probability reads
\be
p_{ij}=\sum_{i'j'}A_{ii'}\rho_{i'j'}B_{j'j}\label{claij}
\ee
where $\rho_{rs}$ is the probability distribution of the systems, $\rho_{i'j'}\geq 0$, $\sum_{i'j'}\rho_{i'j'}=1$,
$A_{ii'}\in[0,1]$ is the probability to read $1$ from $i'$ in the measurement $i\geq 1$ ($A_{0i'}=1$ in the case (a)), and analogously $B_{j'j}$.
By linearity of the determinant with respect to rows/columns and Cauchy-Binet formula, the upper bound turns out to be $4^{-n}$ in the case (a).
In the case (b), the determinant is the product of eigenvalues whose sum is bounded by trace, and the trace is bounded
by $1$ (sum of all probabilities), so by arithmetic and geometric means  the maximum is $(n+1)^{-n-1}$, saturated whenever $d> n$, taking $\rho_{i'}=1/(n+1)$ for $i'=1\dots n+1$ and $A_{ii'}=B_{i'i}=\delta_{ii'}$.

For the limited $d$ in the case (a), the classical extremal cases have been tabularized in Table \ref{tbc}.
In the quantum case of (a), we reach the bound for $n=1,2$ for $d=2$, in the real case, and 
$n=3$ for $d=2$ in the complex case. For  $d=4$ the upper bound $W_n=4^{-n}$ is reached for $n\leq 9$ and $n\leq 15$ in the real and complex cases, respectively. The result for $d=4$ (ququarts) is obtained as follows.
We take two qubits for each party to span $d=4$ space. We use the maximally entangled state $|\psi\rangle=\sum_x|x,x\rangle/2$ with $x=00,01,10,11$ in this space for a single party. In such a case, the probabilities from (\ref{pij}) read 
\begin{align}
&p_{ij}=\sum_{xy}\langle x,x|A_iB_j|y,y\rangle/2=\nonumber\\
&\sum_{xy}(A_i)_{xy}(B_j)_{xy}/4=
\mathrm{Tr}A_iB_j^\ast/4,
\end{align}
 i.e. a direct matrix product (not tensor) for $A_i$ and $B_j$ written in the basis $|x\rangle$,
according to (\ref{pij}). Here the observables are represented by $A_{st}=B^\ast_{st}=(1+\sigma^1_s\sigma^2_t)/2$  with the upper index denoting the qubit, and $st=0,1,2,3$ denoting the standard Pauli matrices. It is clear that any subset of such observables built for pairs $(st)$ (with $st=0,1,3$ except $s=t=0$) will maximize $W$ because of mutual orthogonality in terms of Hilbert-Schmidt scalar product $\mathrm{Tr} XY^\dag$. All observables from the set are real, plus one extra for $s=t=2$. We then get the upper bound for real ququarts up to $n\leq 9$. In the complex case, we can take all possible $(st)$ except $(00)$ to we get $n\leq 15$. For lower $d$ we determined the bounds of $W$ numerically and collected the results in Table \ref{tbq} (see details in Appendix \ref{appb}).

In the case (b), the maximum is classical for $d>n$. For $d\leq n$ the total maximum is $[(d-1)/n]^{n}/(n+1)^{n+1}$, 
saturated for maximally entangled states
and single projection observables $A_i=B_i$ equal single projections corresponding to equiangular tight frames \cite{bb23} times a constant, see details in Appendix \ref{appa}. In cases without equiangular tight frames, one has to determine the maxima numerically. For $n<5$, the only such cases are $n=4$, $d=3$ real $1.6875\cdot 10^{-5}$ and complex $1.8746\cdot 10^{-5}$, lower than the equiangular bound $2\cdot 10^{-5}$, see details in Appendix \ref{appb}.

We also stress that the naive application of a preparation and measurement scenario \cite{opt} 
cannot verify per se the Schmidt number in the bipartite case because of the nonlinearity of the product states (we present a counterexample 
in Appendix \ref{appc}).

\begin{table}
\begin{tabular}{c|*{4}{c}}
\toprule
$n\backslash d$&$2r$&$2c$&$3r$&$3c$\\
\midrule
$1$&$1$&$1$&$1$&$1$\\
$2$&$1$&$1$&$1$&$1$\\
$3$&$0$&$1$&$0.85$&$1$\\
$4$&$0$&$0$&$0.55$&$0.78$\\
$5$&$0$&$0$&$0.38$&$0.69$\\
$6$&$0$&$0$&$0$&$0.54$\\
$7$&$0$&$0$&$0$&$0.35$\\
$8$&$0$&$0$&$0$&$0.25$\\
\bottomrule
\end{tabular}
\caption{The calculated fully quantum maxima for the test (a) of the witness quantity $4^nW_n$ for $d=2(r/c), 3(r/c)$ with $(r/c)$ denoting the real/complex case.}
\label{tbq}
\end{table}

\section{Error analysis.}
To determine the value of the witness $W_n$, we collect data from $N$ repetitions of each measurement combination. 
The uncertainty in determining $W_n$ is analogous to the prepare-measure scheme in Ref.\cite{bb22} for finite statistics and assuming $\langle W_n\rangle=0$. In our notation, the resulting error, $N\Delta W_n^2$, is
\begin{align}
&\sum_{kj}{\mathcal A}^2_{jk}p_{kj}(1-p_{kj})&\mbox{ case (a)},\nonumber\\
&\sum_{kj}{\mathcal A}^2_{jk}p_{kj}-\left(\sum_{kj}{\mathcal A}_{jk}p_{kj}\right)^2&\mbox{ case (b)}\label{err},
\end{align}
with the adjugate matrix $\mathcal A=\mathrm{Adj}\; p$ calculated directly, since $p^{-1}\det p$ does not exist when $\det p=0$.
One should also avoid the situation of $\mathcal A=0$, i.e. when the rank is already smaller, as the error becomes not reliable, and one has to consider 
second-order minors.
In our measure-measure approach, in the case (a), the no-signaling assumption means that, in principle, the probabilities $p_{0j}$ and $p_{k0}$ are not independent. The simplest approach is to take them averaged from other experiments, while still treating them as independent. In this way, we just find an upper estimate for the error. On the other hand, verifying no-signaling is an important sanity check, which we perform in Appendix \ref{appd}.

\begin{figure}
\begin{tikzpicture}[scale=1]
		\begin{yquant*}[register/separation=3mm]
			% q[1];
			qubit {} q[4];
			[name=init]
			init {A $\ket 0$} q[0];
			init {$\ket 0$} q[1,2];
			init {B $\ket 0$} q[3];
			% cbit c[1];
			[name=left]
			box {$S$} q[1];
			[name=right]
			cnot q[2] | q[1];
			[name=leA]
			cnot q[0] | q[1];
			[name=riA]
			cnot q[1] | q[0];
			[name=leB]
			cnot q[3] | q[2];
			[name=riB]
			cnot q[2] | q[3];
			box {$S_{1}$} q[0];	 
			box {$S_{1}$} q[3];
			box {$S_{2}$} q[0];	 
			box {$S_{2}$} q[3];
			measure q[0,3];
		\end{yquant*}
		\node[draw, dashed, fit=(left) (left) (right) (right)] {};
		\node[draw, dotted, fit=(leA) (leA) (riA) (riA)] {};
		\node[draw, dotted, fit=(leB) (leB) (riB) (riB)] {};
\end{tikzpicture}
\bigskip
\bigskip

\begin{tikzpicture}[scale=1]
		\begin{yquant*}[register/separation=3mm]
			% q[1];
			init {A $\ket 0$} q[0];
			init {$\ket 0$} q[1,2,3];
			init {B $\ket 0$} q[4];
			% cbit c[1];
			box {$S$} q[2];
			cnot q[3] | q[2];
			[name=leC]
			cnot q[1] | q[2];
			[name=riC]
			cnot q[2] | q[1];
			cnot q[4] | q[3];
			cnot q[3] | q[4];
			cnot q[0] | q[1];
			cnot q[1] | q[0];
			align -;
			box {$S_{1}$} q[0];	 
			box {$S_{1}$} q[4];
			box {$S_{2}$} q[0];	 
			box {$S_{2}$} q[4];
			measure q[0,4];
		\end{yquant*}
		\node[draw, dotted, fit=(leC) (leC) (riC) (riC)] {};
\end{tikzpicture}
\caption{The implementation of the test (a) on IBM Quantum with two (top) and three (bottom) intermediate qubits.
The dashed box represents entanglement creation while the dotted ones swap entanglement to the next neighbors.
The extra swap box is indicated in the bottom circuit.}
\label{mid23}
\end{figure}
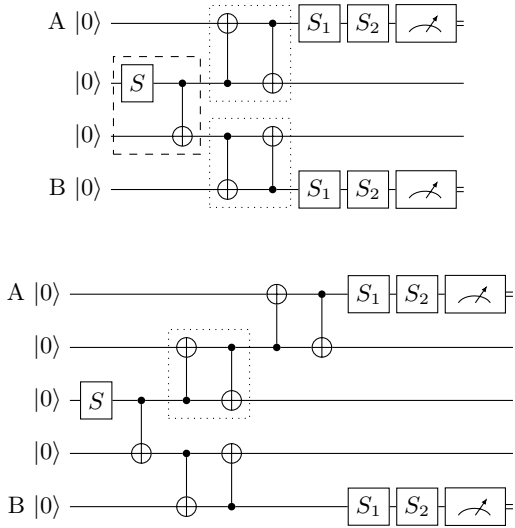

\begin{figure}
\includegraphics[scale=.2]{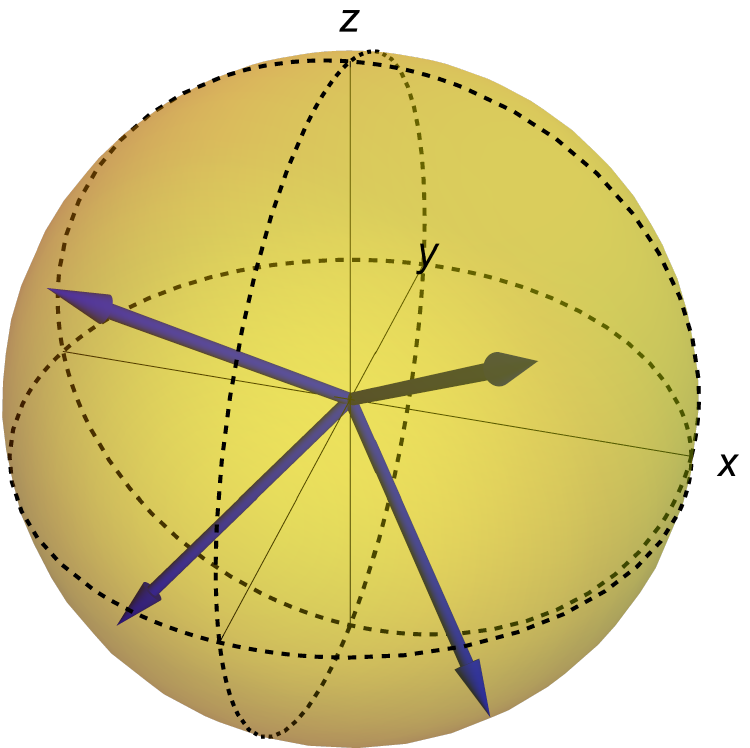}
\includegraphics[scale=.2]{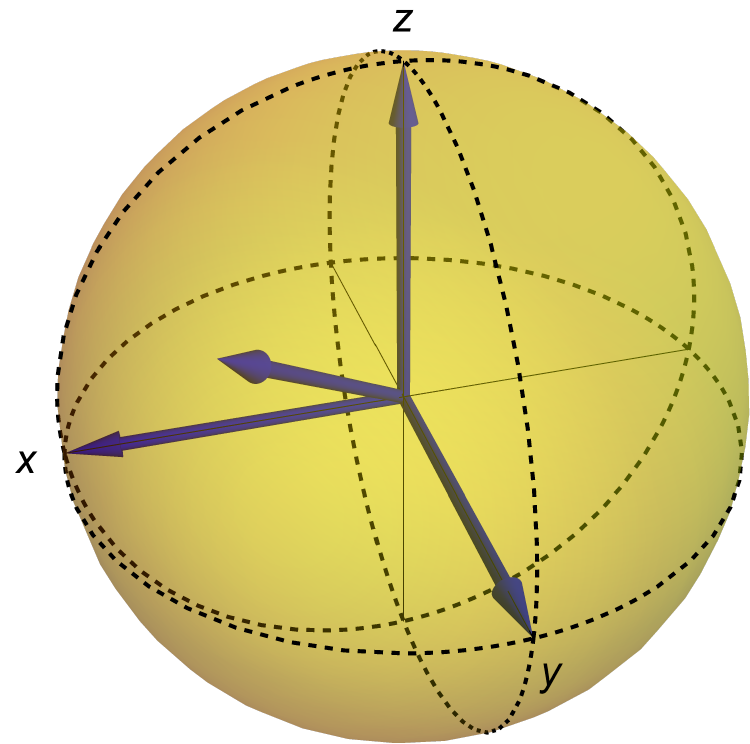}
\caption{The Bloch vectors for the sets of measurements used in the work, left set I and right set II, in the test (a).}
\label{bloch}
\end{figure}

\section{Demonstration on IBM Quantum}
We have demonstrated the feasibility of the above test on IBM Quantum devices.
A microwave pulse tuned to the interlevel drive frequency allows one to apply the parametrically controlled gates. The native single qubit gate is the $\pi/2$ rotation
\begin{equation}
S=\sqrt{X}=(\sigma_0-i\sigma_1)/\sqrt{2}\label{smat},
\end{equation}
in the $|0\rangle$, $|1\rangle$ basis.
The rotation for a given angle $\theta$ is realized with the native gate $S$ and two gates $Z_\theta$
\begin{equation}
S_\theta=Z^\dag_\theta SZ_\theta ,\:
Z_\theta=\sigma_0\cos\theta/2-i\sigma_3\sin\theta/2,
\label{sgate}
\end{equation}
with shorthand notation $Z=Z_{\pi}$, $Z_\pm=Z_{\pm\pi/2}$.
In addition, there is a native two-qubit $CNOT$ gate on most of IBM quantum devices, operating as
\be
|00\rangle\langle 00|+|01\rangle\langle 01|+|11\rangle\langle 10|+|10\rangle\langle 11|,
\ee
where for $|ab\rangle$ the control qubit states  is $a$ (depicted as $\bullet$) and
target qubit state is $b$ (depicted as $\oplus$ in Fig. \ref{mid23}). The newest devices use Echoed Crossed Resonance ($ECR$) gate, instead of $CNOT$ but one can transpile
the latter by additional single-qubits gates, see Appendix \ref{appe}.

In the case (a), we create the Bell state $(|+-\rangle-|-+\rangle)/\sqrt{2}$ with 2 or 3 qubits between $A$ and $B$, where $|+\rangle=|0\rangle$, $|-\rangle=|1\rangle$ for $A$
and $|+\rangle=i|1\rangle$, $|-\rangle=|0\rangle$ for $B$. The final measurement is fixed by two subsequent local gates $S_{1,2}$ from (\ref{sgate}), with appropriately chosen angles $\theta_1, \theta_2$ as explained below.
We took two different sets of measurements: set I and set II. In the language of the Bloch sphere, our measurement directions are
eigenstates of $\vec{a}\cdot\vec{\sigma}$ with the eigenvalue $+1$ for a unit vector $\vec{a}$. We took four vectors for $A$ and the opposite vectors for $B$. 
Vectors in the set I correspond to tetrahedron vertices: 
 $\vec{a}=(\pm 1,\pm 1,1)/\sqrt{3}$. In the set II, $\vec{a}$ are along principal axes $xyz$, and the last direction is $(1,1,1)/\sqrt{3}$,
see Fig. \ref{bloch} for an illustration.
In this way,  we cover the maximal space of qubit states. Both sets have been tested on IBM devices belem qubits $0,4$ (with qubits $1,3$ in the middle)
and lagos, perth, and nairobi $0,6$ (with  qubits $1,3,5$ in the middle). The technical characteristics of the devices are given in 
Appendix \ref{appf}. 

In the case (b), we  create the same Bell state as in the case (a), between next neighbor qubits $A$ and $B$, separated by one extra middle connector qubit,
except that now $|+\rangle=|0\rangle$ and $|-\rangle=|1\rangle$, by additional $Z_-$ and $X=|0\rangle\langle 1|+|1\rangle\langle 0$ (also native) gates at the end.
The separation makes the communication between the parties unlikely, as any reasonable crosstalk, cannot affect the next neighbors.
The 5 outcomes are formally represented by fractions of projections  $M_j=(1+\vec{m}_j\cdot\vec{\sigma})/8$
for the directions $j=1,2,3,4$ in vertices of the rotated tetrahedron
\begin{align}
&\vec{m}_1=(\sqrt{2/3},0, -1/\sqrt{3}),\nonumber\\
&\vec{m}_2=(0,\sqrt{2/3},1/\sqrt{3}),\nonumber\\
&\vec{m}_3=(-\sqrt{2/3},0,-1/\sqrt{3}),\nonumber\\
&\vec{m}_4=(0,-\sqrt{2/3},1/\sqrt{3}),\label{tet}
\end{align}
and $M_5=1/2$ \cite{osz}. The actual implementation requires the following measurement protocol requiring 3 qubits at each party
denoted by $a_0a_1a_2$, $b_0b_1b_2$ for the party $A$, $B$, respectively, with the final results $a_j,b_j=0,1$ 
corresponding to the states $\ket 0,\ket 1$, respectively. Let us focus now on one of the parties, say $A$, as the other party is analogous.
The initially entangled qubit of $A$ is mapped by a $CNOT$ gate to an auxiliary
one, i.e. 
\begin{equation}
\alpha|00\rangle+\beta|10\rangle\to \alpha|00\rangle+\beta|11\rangle
\end{equation}
Now, we choose one of these two qubits to be a spectator qubit, $a_0$, while the other qubit is the
working qubit $a_1$. We apply rotation $Z_+$ to the spectator qubits and subsequently $S$, 
to measure them in the $x$ basis i.e. $(|0\rangle\pm |1\rangle)/\sqrt{2}$. 
In this way, the probability $p(a_0)$ is always $1/2$. For $a_0=0,1$, the working qubit  is in the state $\alpha|0\rangle+\beta|1\rangle$
or $\alpha|0\rangle-\beta|1\rangle$, respectively. Note that the latter state differs from the first one by $Z_\pi$ rotation.
We finally apply $Q$ gate to map the projections (\ref{tet}) of each working qubit onto the four states of the two qubits $a_1a_2=00,10,01,11$, corresponding to outcomes $1,2,3,4$, see details in Appendix \ref{appe} and Fig. \ref{qgate}.
Note that the $Z$ rotation on $Q$ projection simply reverses the bit $a_2$.
The original set of 8 outcomes can be treated as 5 outcomes, depending on the spectator qubit,
i.e. a chosen value $a_0$ defines 4 outcomes $a_0a_1a_2$ and the rest $(1-a_0)\ast\ast$ are treated as 5 (i.e. 4 other outcomes are in a single set).
Out of all 64 probabilities $p_{ab}$, 16 of them are zero, for $a_1=b_1$ and $a_0+a_2-b_0-b_2$ even. The remaining probabilities are each equal $1/48$.
The complete scheme, using $CNOT$ gate is depicted in Fig. \ref{mid55}.

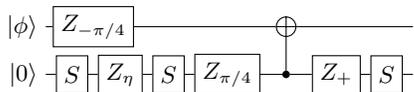
\begin{figure}
\begin{tikzpicture}[scale=1]
		\begin{yquant}
			qubit {} q[2];
			[name=init]
			init {$\ket 0$} q[1];
			init {$\ket \phi$} q[0];
			box {$S$} q[1];
			box {$Z_\eta$} q[1];
			box {$S$} q[1];
			box {$Z_{\pi/4}$} q[1];
			box {$Z_{-\pi/4}$} q[0];
			cnot  q[0] | q[1];
			box {$Z_+$} q[1];
			box {$S$} q[1];
		\end{yquant}
\end{tikzpicture}

\caption{The $Q$ gate mapping the working state $|\phi\rangle$ onto the 4 projections (\ref{tet}) for the measurements $00,10,01,11$. We used the angle $\eta=\mathrm{acos}\sqrt{1/3}$}
\label{qgate}
\end{figure}

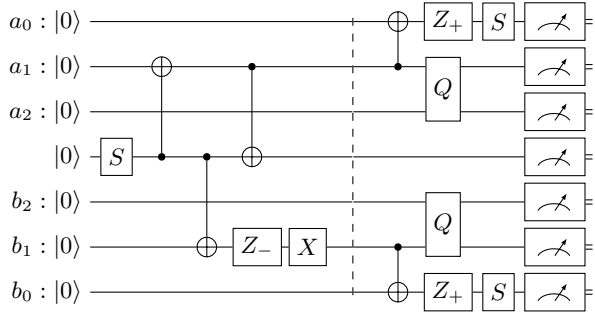
\begin{figure}
\begin{tikzpicture}[scale=1]
		\begin{yquant*}
			% q[1];
			init {$a_0:\ket 0$} q[0];
			init {$a_1:\ket 0$} q[1];
			init {$a_2:\ket 0$} q[2];
			init {$\ket 0$} q[3];
			init {$b_0:\ket 0$} q[6];
			init {$b_1:\ket 0$} q[5];
			init {$b_2:\ket 0$} q[4];
			% cbit c[1];
			box {$S$} q[3];
			cnot q[1] | q[3];
			cnot q[5] | q[3];
			cnot q[3] | q[1];
			%cnot q[1] | q[0];
			box {$Z_-$} q[5];
			box {$X$} q[5];	
			barrier (q);
			cnot q[0] | q[1];
			cnot q[6] | q[5];
			box {$Z_+$} q[0];
			box {$Z_+$} q[6];
			box {$S$} q[0];
			box {$S$} q[6];
			box {$Q$}  (q[1,2]);

			box {$Q$}  (q[4,5]);
			measure q[0-6];
		\end{yquant*}
\end{tikzpicture}

\caption{The test (b) using $CNOT$ and $Q$ gates (see Fig. \ref{qgate}) using 7 qubits, in two groups of 3 qubits, party $A$: $a_0a_1a_2$, party $B$: $b_0b_1b_2$ , separated by an extra connector qubit. }
\label{mid55}
\end{figure}

Each test consists of a certain number of jobs, where each circuit (randomly shuffled) is run
a certain number of shots. Since the number of experiments is 16, each one could be repeated to saturate the limit on circuits. The total number of trials is 
$N=\#jobs\#shots\#repetitions$.  Due to calibration changes, every several hours, the probabilities may drift, which can affect the witness being 
a nonlinear function of probabilities. To take it into account, we have calculated the witness in two ways \cite{ibm}: $W$ and its error is obtained from total probabilities of all jobs together, $W'$ and its error is obtained by calculating $W$ and the error for each job individually, and then averaging it over jobs.
It turns out that these values indeed differ but do not change the verdict about Schmidt number.

\begin{figure}
\includegraphics[scale=.9]{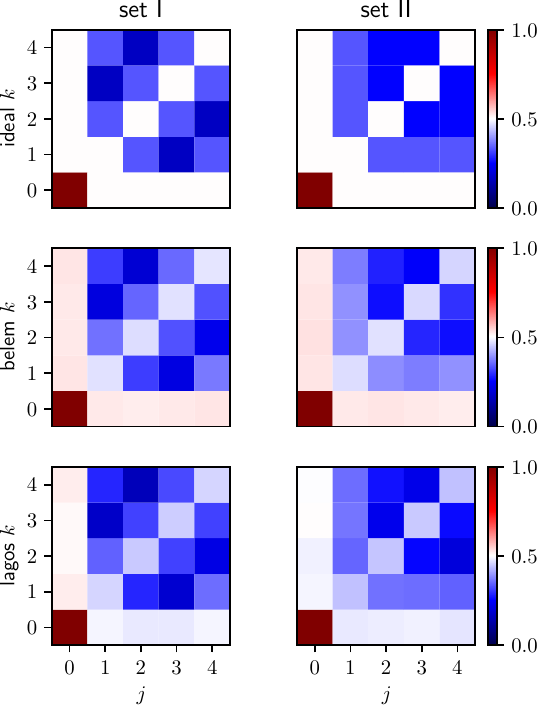}
\caption{Results of the tests (a)  with probabilities $p_{kj}$, for belem and lagos, compared to the ideal expectation.}
\label{mesh}
\end{figure}

\begin{figure}
\includegraphics[scale=.8]{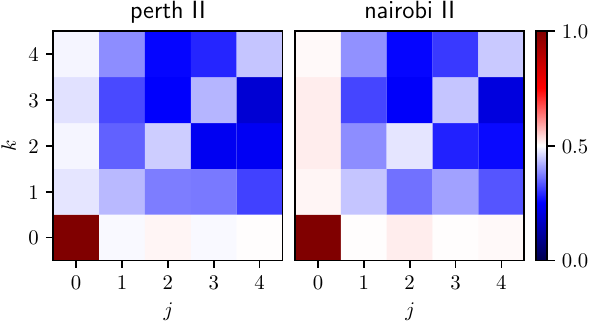}
\caption{Results of the tests (a) with probabilities $p_{kj}$, for perth and nairobi}
\label{meshp}
\end{figure}

For the case (a),
we run $247/404$ jobs on belem with $20000$ shots and $6$ repetitions
for sets I/II,
while $20/15$ jobs with $10000$ shots and $6$ repetitions on lagos. An additional test has been run on perth/nairobi (set II)
with $259/550$ jobs, $10000$ shots and $6$ repetitions.
The statistics differ from the ideal ones because of the noise but agree qualitatively, see Fig. \ref{mesh} and \ref{meshp}.
The final witness agrees with the null hypothesis for $d=2$ within the statistical error, see Table \ref{resu}. No-signaling has been confirmed as a sanity check.
However, in the test on perth, we found the desired value of the witness,
but we also observed a moderate violation of no-signaling at the level of $3.9$ standard deviation, 
see Appendix \ref{appd}. Due to $48$ possible comparisons, the look-elsewhere-effect lowers the significance of this observation.
Nevertheless, this difference suggests that our test may be useful in identifying malfunctions of the devices in the future.

\begin{table}
\begin{tabular}{*{5}{c}}
\toprule
device &$W_4$&$\Delta W_4$ & $W'_4$ & $\Delta W'_4$\\
\midrule
belem I&$0.199$&$1.267$ & $0.135$ & $1.294$\\
lagos I&$15.69$&$9.08$ & $20.52$ & $9.75$\\
belem II&$-0.647$&$1.269$ & $-3.749$ & $1.338$\\
lagos II&$-8.233$&$9.782$ & $-7.402$ & $9.992$\\
perth II&$1.664$&$2.132$ & $1.551$ & $2.139$\\
nairobi II&$2.184$&$1.448$ & $2.293$ & $1.483$\\
\bottomrule
\end{tabular}
\caption{The values of the witnesses for the test (a) with errors depending on the device and set I/II. Here $W_4$, $W'_4$, errors $\Delta W_4$, and $\Delta W'_4$ are in units $10^{-6}$.}
\label{resu}
\end{table}

\begin{figure}
	\includegraphics[scale=.8]{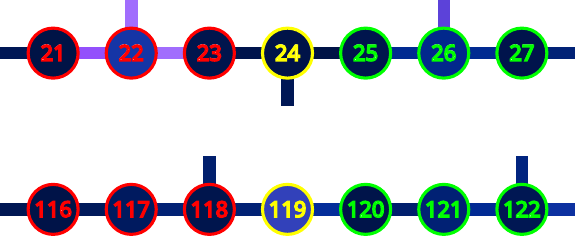}
	\caption{The tested groups of qubits on IBM Quantum brisbane. We have highlighted
	    qubits in the test (b), parties A (red), B (green), and middle connectors (yellow).
		Two-qubit Echoed Cross Resonance gates, used in the test, connect the
		tested qubits, and the external ones.}
	\label{brism}
\end{figure}

\begin{table}
	\begin{tabular}{*{8}{c}}
		\toprule
		group&  $a_0b_0$  &  $W_4$ &$\Delta W_4$&  $W'_4$& $\Delta W'_4$\\
		\midrule
		nairobi 	&00&46.9& 7.5 &56.1 & 8.2\\
		0,1,2 (A)	&01&-42.4& 6.6 &-50.3 & 7.2\\
		6,5,4 (B)	&10&-48.7& 7.1 &-54.5 & 7.6\\
					&11&42.6& 6.3 &50.7&6.7\\
		\midrule
		brisbane 			&00&-69.9& 8.7 &-69.3 &9.0\\
		23,22,21 (A) 		&01&68.8& 9.8 &  67.7 & 10.1\\ 
		25,26,27 (B)	    &10&64.4& 8.5 & 65.8 & 8.9\\ 
							&11&-62.0& 9.4 & -63.0 & 9.9\\
		\midrule
		brisbane 			&00&-30.0& 3.5 & -32.4 & 3.8\\
		118,117,116 (A) 	&01&29.5& 3.4 & -31.7 & 3.7\\
		120,121,122 (B)	    &10&28.7& 3.3 & 30.9 & 3.6\\
							&11&-29.1& 3.4 & -31.4 & 3.7\\ 
		\bottomrule
	\end{tabular}
	\caption{Experimental results from the test (b) for nairobi and the two groups on brisbane, for all combinations of spectator qubit values $a_0b_0$.
Here $W_4$, $W'_4$, errors $\Delta W_4$, and $\Delta W'_4$ are in units $10^{-12}$.}
	\label{ress}
\end{table}

The test (b) has been performed on nairobi, qubits 0,1,2 (A) and 6,5,4 (B) connected by 3, and brisbane, 2 groups, qubits 23,22,21 (A) and 25,26,27 (B), connected 
by 24, and 118,117,116 (A) and 120,121,122 (B) connected by 119, as depicted in Fig. \ref{brism}. For nairobi, we have run 20 jobs, with 300 repetitions and 100000 
shots. For brisbane, we have run 20 jobs, 75 repetitions, and 20000 shots. The results are presented in the Table \ref{ress}. The results show the failure of the order of
6, 7, and 8 standard deviations. 
To explain the nonzero value of $W$ by small corrections to probabilities of any origin, i.e. $p\to p+\delta p$, one can estimate
$
\delta W=\mathrm{Tr} \mathcal A\delta p$ for the adjugate matrix $\mathcal A$.
The elements of the adjugate matrix are of the order $\lesssim 10^{-6}$ for nairobi and $\lesssim 10^{-7}$ for brisbane,
so the result of the order $\gtrsim 10^{-11}$ would require the contribution from an extra state, beyond the assumed Hilbert space, of the order $10^{-5}$ or $10^{-4}$. Taking into account
that the entanglement is created by a single $S$ gate with error $\lesssim 10^{-3}$ (includes leakage to higher excited states) and two $CNOT/ECR$ gates with errors $\sim 10^{-2}$ (the error includes known crosstalks), the total 
technical contribution would be $<10^{-7}$  (product of error of $S$ and two $CNOT/ECR$s to reach the parties $A$ and $B$) which is much below our observation. 
We have checked also that the values of the witness for individual jobs are consistently nonzero and their average is close to the total values (\ref{ress}).
The diagnostic data are summarized in Appendix \ref{appf} and the raw data are available publicly \cite{zen}.

\section{Conclusions}
We have demonstrated the extreme usefulness of null  tests of the Schmidt number for the bipartite states, which should help in the diagnostics of quantum
devices. It is complementary to the violation of a Bell-type inequality while based on similar assumptions.
Combining it with no-signaling verification and Bell-type violations in a single test can serve as a powerful quality criterion
of multiqubit networks. We stress that our null hypothesis tests remain robust against most common disturbances, as long as they are local,
with known mechanisms. Due to the extreme accuracy of the test, we were able to diagnose IBM Quantum devices, far beyond standard technical specifications. The 
results showed consistency with the Schmidt number $d=2$ in the case of independent measurements 
but the test with many outcomes failed to confirm it.
The deviation is significant and exceeds possible common origins due to gate errors. The failure requires an urgent technical explanation.
Otherwise, the results may be a signature of an exotic picture, involving e.g. many worlds/copies \cite{plaga,abadp}
($N$ copies of the same system formally boost the dimension and Schmidt number from $d$ to $d^N$). We refrain from giving an exact model, as the collected data are
insufficient to draw stronger conclusions.

\section*{Acknowledgments} The results have been created using IBM Quantum. The views expressed are those of the authors and do not reflect the official policy or position of IBM Quantum team. 
The project is co-financed by the funds of the Polish
Ministry of Education and Science under the program entitled International
Co-Financed Projects. TR gratefully acknowledges the funding support by the
program ,,Excellence initiative research university'' for the AGH University in
Krakow as well as the ARTIQ project: UMO-2021/01/2/ST6/00004 and
ARTIQ/0004/2021.

\appendix

\section{Details of the proof of the bound on $W_n$ in the classical and quantum case}
\label{appa}

Here we present the proof of the bound on $W_n$ over possible states and measurements.
In the classical case (a), the maximization relies on basic linear algebra and the Cauchy-Binet formula for the determinant of a product of rectangular matrices.
Taking two elements of $\rho$ in a single column/row, e.g. $(1,\pm)$ and 
shifting $\rho_{1\pm}\to \rho_{1\pm}\pm\Delta$ results in $p_{ij}\to p_{ij}+\sum_{\pm} p_{ij}(\pm) A_{i1}B_{\pm j}\Delta$.
It implies adding linearly dependent columns/rows, so by decomposing linearly columns in the new determinant to old $p$ and $\Delta$ entries,
only terms linear in $\Delta$ are added.  It means that the determinant is linear in terms of $\Delta$
and its maximum is at extremal $\Delta$, meaning that one of $\rho_{1\pm}$ must be $0$. Repeating this reasoning we end up with the diagonal $\rho$, shortening $\rho_z\equiv \rho_{zz}$.
From linearity with respect to rows/columns, the maximum of $\det p$ occurs when the entries of matrices $A$ and $B$ are either 0 or 1.

From the Cauchy-Binet formula
\begin{align}
&\det p=\sum_{M}\det A_{M}\det B_{M}\prod_{z\in M}\rho_z\leq\nonumber\\
& \sum_{M}(\det A_{M}^2+\det B_M^2)\prod_{z\in M}\rho_z/2\label{cabi}
\end{align}
where $M$ is a subset of indices to restrict columns in $A$ and rows in $B$, and we use the Cauchy inequality at the end.
Therefore to maximize $\det p$ we should replace $A$ with $B^T$ or vice versa, to obtain $p=A\rho A^T$.
Then we end up with a symmetric matrix $p$. Let us now multiply the all  rows  of $A$ except the zeroth one by $2$, which makes $\det p$ multiplied by $4^n$.
Now subtract it from all other rows, which does not change $\det p$ but  the new matrix $A$ consists of $\pm 1$ entries.
The diagonal elements are then $1$  from normalization, so $\mathrm{Tr}\; p=n+1$. For the symmetric matrix the $\mathrm{Tr}\; p=\sum_i\lambda_i$, for eigenvalues $\lambda_i$, and
$\det p=\prod_i\lambda_i$. From inequality between arithmetic and geometric means, the determinant is 
maximal is when all eigenvalues are equal $1$, giving here $1$, and finally $4^{-n}$ for the original determinant.
The maximum is obtained when the rows of the new $A$ are orthogonal. It suffices to have $2^n$ states in $\rho$ (columns of $A$) 
and fill the first $n$ rows in each column by its
binary digits. Then our auxiliary matrix $A$ with $\pm 1$ entires has mutually orthogonal rows, giving the diagonal $p$, proportional to identity, with $\rho_i=2^{-n}$.
In fact, the number of required states is the size of the nearest Hadamard matrix (i.e. $\pm 1$ matrix of orthogonal rows), which has a size often much smaller than $2^n$.

To find the quantum maximum  in both cases, let us begin with the pure state with Schmidt decomposition,
\be
|\psi\rangle=\sum_k \psi_k|kk\rangle
\ee
for real $\psi_k\geq 0$, with $\sum_k \psi^2_k=1$.

Now, we decompose $A=A_D+A_R+iA_I$ with real diagonal $A_D$, and off-diagonal $A_R$ and $A_I$ ($A_I=0$ in the quantum real case), and similarly $B$.
From Hermicity $A_R$ is symmetric and $A_I$ is antisymmetric. 
The matrix $p$ can be expressed 
\begin{align}
&p_{ij}=\sum_{kl}\psi_k\psi_l(A_{i})_{kl}(B_{j})_{kl}=\nonumber\\
&\sum_k \psi_k^2 A_{Dk}B_{Dk}+\sum_{k<l}\psi_k\psi_l(A_{Rkl}B_{Rkl}-A_{Ikl}B_{Ikl}).
\end{align}
Now, we can use Cauchy-Binet formula analogous to (\ref{cabi}), with $z$ running over $Dk$, $Rkl$, and $Ikl$, just like in the classical case.
We also maximize the determinant by replacing either $A$ by $B^T=B^\ast$ or vice versa, noting the minus sign at $I$ product.

The quantum maximum cannot exceed the classical one in the case (a). We can consider only projections $A_i$ as
the Cauchy-Binet formula is quadratic and convex with respect to each individual $A_i$, a linear combination of projections, and so
the maximum requires extremal arguments.
Now subtracting half of the last row and column from the other ones,
and then multiplying each row and column except by 2 the last one, we replace projections $A$ by $2A-1$, which are observables
with outcomes $\pm 1$. The matrix is symmetric since we have a Schmidt-type entangled state and $A=B^\ast$ while each diagonal term is nonnegative and bounded by $1$. Analogously as in the classical case, the bound on the determinant is $1$,
divided finally by $4^n$ because of doubling the values of rows and columns.

In the  case (b), the maximum is classical for $d>n$. Otherwise, 
we can construct the diagonal matrix $\Psi=\mathrm{diag}(\sqrt{\psi_1},\dots,\sqrt{\psi_d})$.
so that $p_{ij}=\mathrm{Tr} A'_i A^{\prime T}_j$ for $A'_i=\Psi A_i\Psi$. 
Let $a_i=\mathrm{Tr}A'_i$, and $A'_i=a_i\bar{A}_i$, so that $\mathrm{Tr}\bar{A}_i=1$. Then $\sum_i a_i=\sum_i \psi_i=c$. Under this constraint, by convexity of Cauchy-Binet expansion with respect to $A'$ we retrieve the  maximal result when $\bar{A}_i$ are single-dimensional projections. Moreover, writing
$\bar{p}_{ij}=\mathrm{Tr}\bar{A}_i\bar{A}^T_j$ we have $\det p=\det\bar{p}\prod_i a^2_i$. From  the arithmetic and geometric mean inequality, the maximum occurs when $a_i=c/n$.
The problem of maximal $\det\bar{p}$ reduces now to our previous result \cite{bb23}, i.e. the maximum of $\det\bar{p}$ is $(n+1)^{n+1}(d-1)^{n}/d^{n+1}n^n$. 
Finally we find the maximal $c=d^{1/2}$ when all $\psi_i=d^{-1/2}$ from the inequality between quadratic and arithmetic means, which completes the final bound.

\section{Determination of classical and quantum maxima in particular cases}
\label{appb}

To find the respective maxima, we have used a hybrid approach similar to \cite{bb22}.
Setting the space of parameters of measurements and states, we find first the maximum numerically.
Then we try to find a symmetry to represent the case using fewer parameters.

\subsection{ Case (a) $n$ independent measurements}

Case $n=1$. $W=0$ for $d=1$ (classical, quantum real or complex).
For $d=2$ we reach the classical maximum for $\rho_j=1/2$ and  (omitting the last row of 1s)
\be
A=\begin{pmatrix}
1&0
\end{pmatrix}
\ee giving $\det=1/4$.

Case $n=2$. $W=0$ for $d=1,2$ classical while $1/3^3$ for $d=3$, with $\rho_j=1/3$, 
\be
A=\begin{pmatrix}
1&0&0\\
0&1&0
\end{pmatrix}.
\ee
The absolute maximum (from Sec. II) is for $d=4$, $\rho_j=1/4$ and
\be
A=\begin{pmatrix}
1&1&0&0\\
1&0&1&0
\end{pmatrix}.
\ee

Case $n=3$. $W=0$ for $d=1,2,3$ classical. The absolute maximum  is for $d=4$, $\rho_j=1/4$, and
\be
A=\begin{pmatrix}
1&1&0&0\\
1&0&1&0\\
1&0&0&1
\end{pmatrix}.
\ee
Case $n=4$. $W=0$ for $d\leq 4$ classical. For $d=5$, $\rho_j=1/5$, and
\be
A=\begin{pmatrix}
1&1&0&0&0\\
1&0&1&0&0\\
1&0&0&1&0\\
1&0&0&0&1
\end{pmatrix},
\ee
we get $W=9/5^5$.

For $d=6$,
\be
A=\begin{pmatrix}
1&  1& 0& 0& 0 &1\\
1&  0& 1& 0& 0 &1\\
0&  0& 0& 0& 1 &1\\
0&  1& 1& 0& 0 &1
\end{pmatrix},
\ee
and $\rho=\mathrm{diag}(x, x, x, z, y, y)$
with $z=1 - 3 x - 2 y$, gives $W=
0.002954143422708182$, $x= 0.19585843826556898$, 
 $y=0.18219100818175962$.
 
For $d=7$,
\be
A=
\begin{pmatrix}
 0& 0& 0& 0& 0& 1& 1\\
 0& 0& 0& 1& 1& 1& 0\\
 0& 0& 1& 1& 0& 0& 1\\
 0& 1& 0& 0& 1& 0& 1
\end{pmatrix},
\ee
with $\rho=\mathrm{diag}(w, x, x, y, y,z,z)$ and $w=1-2x-2y-2z$.
The numerical maximum is $W=0.0030764392399879$ for $x=
0.06135153414853146$, $y=0.1710023907787869$, $z=
0.19069830365543322$.

The absolute maximum occurs for $d=8$ and
\be
A=\begin{pmatrix}
0&1&0&1&1&0&0&1\\
0&0&0&1&0&1&1&1\\
0&1&1&0&0&0&1&1\\
0&0&1&0&1&1&0&1
\end{pmatrix},
\ee
with $\rho_j=1/8$.

Case $n=5$.
For $d=6$ and 
\be
A=\begin{pmatrix}
1&0&1&0&0&0\\
0&1&1&1&0&1\\
1&1&0&0&0&1\\
1&0&0&1&0&1\\
0&0&1&0&1&1
\end{pmatrix},
\ee
and $\rho_j=1/6$
we get $W=5^2/6^6$.

For $d=7$ we get $W=12^{-3}$ for
\be
A=\begin{pmatrix}
0&1&  0& 0& 0& 1&  1\\
0&1&  0& 1& 1& 0&  0\\
0&0&  1& 1& 0& 1&  0\\
0&0&  1& 0& 1& 0&  1\\
0&1&  1& 0& 0& 0&  0
\end{pmatrix},
\ee
and $\rho=\mathrm{diag}(1/6,1/6,1/6,1/8,1/8,1/8,1/8)$.

The absolute maximum occurs for $d=8$, $\rho_j=1/8$ and
\be
A=\begin{pmatrix}
0&0&0&0&1&1&1&1\\
0&0&1&1&0&1&1&0\\
0&1&1&0&1&0&1&0\\
1&0&1&0&0&0&1&1\\
1&0&1&0&1&1&0&0
\end{pmatrix}.
\ee

Quantum case $n=2$, $d=2$.
Bell state
$|\psi\rangle=(|12\rangle-|21\rangle)/\sqrt{2}$
with $A_j=B_j=|v_j\rangle\langle v_j|$
and $|v_1\rangle=|1\rangle$, $v_2\rangle=(|1\rangle+|2\rangle)/\sqrt{2}$
gives the absolute maximum.

Quantum case $n=3$, $d=2$, zero in the real case. Complex: the same Bell state but adding
$
|v_3\rangle=(|1\rangle+i|2\rangle)/\sqrt{2}
$
gives the absolute maximum up to a sign, depending on ordering.

Quantum case $n=d=3$ (real). The state reads
\be
|\psi\rangle=p|11\rangle+q(|22\rangle+|33\rangle),
\ee
with $A_j=B_j=|v_j\rangle\langle v_j|$,
and 
\be
|v_j\rangle=s|1\rangle+w(\cos(2j\pi/3)|2\rangle+\sin(2j\pi/3)|3\rangle),
\ee
constrained by $p^2+2q^2=s^2+w^2=1$
which gives the maximal $W=0.013208219549514474$
for $q=0.5080857929626221$,  $s=0.7236153449503123$.

Quantum case $n=4$, $d=3$ real.
We take
\be|\psi\rangle=(2|11\rangle+\sqrt{3}(|22\rangle+|33\rangle))/\sqrt{10}, 
\ee
and $A_j=B_j=|v_j\rangle\langle v_j|$,
with
\be
v_{1,2}=s_1|1\rangle\pm w_1|2\rangle,\;v_{3,4}=s_2|1\rangle\pm w_2|3\rangle,
\ee
and $s^2_{1,2}=(9\pm\sqrt{17})/16$, $w^2_{1,2}=(7\pm\sqrt{17})/16$, giving $W=27/12500$.
In the complex case 
\be|\psi\rangle=p|11\rangle+q|22\rangle+r|33\rangle,
\ee
and 
\be|v_j\rangle=a|1\rangle+ x\omega^j|2\rangle+y\omega^j|3\rangle,
\ee
for $j=1,2,3$ and $\omega=e^{2\pi i/3}=(i\sqrt{3}-1)/2$
while $|v_4\rangle=b|2\rangle+c|3\rangle$.
Maximizing with constraints $p^2+q^2+r^2=a^2+x^2+y^2=b^2+c^2=1$
we get $\det=0.003065301182016068$
for $x=-0.20660676061609246$, $y=0.8141407994847997$,
$b=0.5366502440643837$, $c=0.8438048656782298$,
$q=0.45755959305674204$, $r=0.6898510489488422$.

Quantum case $n=5$, $d=3$ real.
Then
$|\psi\rangle=(|11\rangle+|22\rangle+|33\rangle)/\sqrt{3}$
and
$$
|v_j\rangle=a(\cos(2\pi j/5)|1\rangle+\sin(2\pi j/5)|2\rangle)+b|3\rangle
$$
with $a^2+b^2=1$. We get the maximum 
$W=(2437+340\sqrt{10})/(2\cdot 3^{14})\simeq
0.0003671542094938571$ for $a^2=(10+\sqrt{10})/15$.

Complex case. We take
$|\psi\rangle=p|11\rangle+q|22\rangle+r|33\rangle$
and $|v_j\rangle=x\zeta^j|1\rangle+y|2\rangle+z\zeta^{-j}|3\rangle$
with $\zeta=e^{2\pi i/5}$. Maximizing with constraints $p^2+q^2+r^2=x^2+y^2+z^2=1$
we get $W=0.000674047929103352$, $x=0.7998181925131095$, $z=-0.4434461617437569$, 
$p=0.6838826680323404$, $r=0.5298910387696789$.

Quantum $n=6$, $d=3$ complex. The state
$|\psi\rangle=\sqrt{2/7}(|11\rangle+|22\rangle)+\sqrt{3/7}|33\rangle$
and $|v_j\rangle=(|1\rangle+\omega^j|3\rangle)/\sqrt{2}$, $|v_{j+3}\rangle=(|2\rangle+\sqrt{2}\omega^j|3\rangle)/\sqrt{3}$
returns $W=4\cdot 27/7^7$.

Quantum $n=7$, $d=3$ complex.
We take $|\psi\rangle=\sqrt{5}(|11\rangle+|22\rangle)/4+\sqrt{6}|33\rangle/4+$ and
$|v_{j+3m}\rangle=x_m|1\rangle+y_m|2\rangle+z_m \omega^j|3\rangle$ for $j=1,2,3$, $m=0,1$
and $|v_7\rangle=|1\rangle$
Then the numerical maximum with constraints $x_m^2+y_m^2+z_m^2=1$ is $W=0.0000215113826$.

Quantum $n=8$, $d=3$ complex.
$|\psi\rangle=(|11\rangle+|22\rangle+|33\rangle)/\sqrt{3}$
with 
$|v_j\rangle=\sqrt{10}|1\rangle/6\pm i|2\rangle/\sqrt{6}+\sqrt{5}\omega^j|3\rangle/3$
for $j=1\cdots 6$ while $|v_{7,8}\rangle=\sqrt{5/6}|1\rangle\pm |2\rangle/\sqrt{6}$
which gives $W=5^{10}/3^{26}$.

\subsection{Case (b) single measurement with $n+1$ outcomes}

It is convenient to consider positive Hermitian matrices $A'_i=\sqrt{Z}\Psi A_i\Psi$, which are not normalized, giving $p'_{ij}=\mathrm{Tr} A'_iA'_j$
and normalize probability by $p=p'/Z$ with $Z=\mathrm{Tr} A^2$ for $A=\sum_i A'_i$. The state is then reconstructed as $\Psi^2=A/\sqrt{Z}$ in its diagonal basis.
Similarly as in \cite{bb23} we focus on matrices of rank $\leq d/2$. Otherwise, the overlapping projection is free to shift between the matrices
with  the linear change of the determinant. In the case $n=4$, $d=3$ we could reduce $i<5$ to rank 1. The last matrix $i=5$ could have rank 2, but
our numerical analysis showed that the optimal case is also of rank 1. The result in the real case
$A'_j=|v_j\rangle\langle v_j|$ with
\begin{align}
&|v_{1,2}\rangle=x|1\rangle\pm y|2\rangle,\nonumber\\
&|v_{3,4}\rangle=a|1\rangle\pm b|3\rangle,\nonumber\\
&|v_5\rangle=z|2\rangle,
\end{align}
with $a^4=1/160$, $b^4=1/120$, $y=3b$, $x=4a$, $z=12b$ giving $\det=1.6875\cdot 10^{-5}$ (exact)

In the complex case,
$A'_j=|v_j\rangle\langle v_j|/Z$ with 
\begin{align}
&|v_j\rangle=a|j\rangle,\:j=1,2,\nonumber\\
&|v_j\rangle=q\omega^j|1\rangle+q\omega^{2j}|2\rangle+r|3\rangle,\:j=3,4,5,
\end{align}
with $\omega=e^{2\pi i/3}=(\sqrt{3}i-1)/2$, $a^2=3q^2(q^2+2r^2)/(q^2+r^2)$ and $Z=9(2q^4(2q^2+3r^2)^2+r^4(q^2+r^2)^2)/(q^2+r^2)$.
As the determinant is a homogenous function of $q$ and $r$, setting $q=1$ and $r^2=x$ we end up with
\begin{equation}
\det=\frac{x^2 (1 + x)^6 (1 + 2 x)^6)}{27 (x^2 (1 + x)^2 + 2 (2 + 3 x)^2)^5},
\end{equation}
reaching its maximum $1.874577768244\cdot 10^{-5}$ for $x$ being the largest root of
\begin{equation}
x^4+2x^3-11x^2-11x-2=0
\end{equation}
i.e. $x=2.98813453198126056781$.

\begin{table}
\begin{tabular}{*{4}{c}}
\toprule
device/qubit & freq. (GHz)& anh. (GHz)&$S$ error $[10^{-4}]$\\
\midrule
belem 0&5.09&-0.34&$1.8$\\
belem 1&5.25&-0.32&$6$\\
belem 3&5.17&-0.33&$2.3$\\
belem 4&5.26&-0.33&$2.3$\\
lagos 0&5.24&-0.34&$3$\\
lagos 1&5.10&-0.34&$1.8$\\
lagos 3&4.99&-0.345&$2.3$\\
lagos 5&5.18&-0.34&$1.9$\\
lagos 6&5.06&-0.34&$1.9$\\
perth 0&5.16&-0.34&$3$\\
perth 1&5.03&-0.34&$2.5$\\
perth 3&5.13&-0.345&$2.2$\\
perth 5&4.98&-0.34&$2.8$\\
perth 6&5.16&-0.34&$2.9$\\
nairobi 0&5.26&-0.34&$2.8$\\
nairobi 1&5.17&-0.34&$3.4$\\
nairobi 2&5.27&-0.34&$3.8$\\
nairobi 3&5.03&-0.34&$4.2$\\
nairobi 4&5.18&-0.34&$2.5$\\
nairobi 5&5.29&-0.34&$3.9$\\
nairobi 6&5.13&-0.34&$1.6$\\
brisbane 21&4.97&0.31&$2.1$\\
brisbane 22&5.04&0.31&$3.9$\\
brisbane 23&4.84&0.31&$2.1$\\
brisbane 24&5.01&0.31&$2.7$\\
brisbane 25&4.95&0.31&$3.5$\\
brisbane 26&4.85&0.31&$2.7$\\
brisbane 27&4.75&0.31&$1.4$\\
brisbane 116&4.91&0.31&$1.3$\\
brisbane 117&4.83&0.31&$4.0$\\
brisbane 118&4.73&0.31&$1.4$\\
brisbane 119&4.80&0.31&$2.7$\\
brisbane 120&4.84&0.31&$3.1$\\
brisbane 121&4.97&0.31&$3.6$\\
brisbane 122&4.94&0.31&$3.0$\\
\bottomrule
\end{tabular}
\caption{The characteristics of the single qubits used in the demonstration,
frequency between 0 and 1 level, anharmonicity (frequency between 1 and 2 levels above 0-1 transition), error of the gate $S=\sqrt{X}$ used in the tests. The duration of the single gate pulse is always 35ns, except brisbane with 60ns.}
\label{tech}
\end{table}
\begin{table}[t!]
\begin{tabular}{*{3}{c}}
\toprule
device&connection &$CNOT/ECR$ error $[10^{-3}]$\\
\midrule
nairobi &0-1&$9.2$\\
nairobi &1-3&$6.5$\\
nairobi &1-2&$8.5$\\
nairobi &3-5&$16.1$\\
nairobi &4-5&$4.8$\\
nairobi &5-6&$6.1$\\
brisbane &21-22&$5.2$\\
brisbane &22-23&$7.1$\\
brisbane &24-23&$6.6$\\
brisbane &25-24&$7.1$\\
brisbane &26-25&$10.1$\\
brisbane &27-26&$6.1$\\
brisbane &116-117&$6.1$\\
brisbane &117-118&$5.7$\\
brisbane &118-119&$11.5$\\
brisbane &120-119&$20.5$\\
brisbane &121-120&$7.3$\\
brisbane &122-121&$10.4$\\
\bottomrule
\end{tabular}
\caption{The errors of the two-qubit gates used in the demonstration, $CNOT$ for nairobi and $ECR$ for brisbane.}
\label{tech2}
\end{table}

\section{Counterexample to prepare and measure scenario}
\label{appc}

 In \cite{opt}, one considers  $W_7=\det p$ for the matrix $p$
 with entries $p_{ij}=p(i|2j)-p(i|2j+1)$ for $p(k|i)$ for $i=0\dots 6$, and $j=0\dots 13$ being the probability 
 of $1$ the $i$th dichotomic  measurement on the state $k$ \cite{dim}.
 It is claimed that $W_7=0$ in the case of a bipartite state consisting of two independent qubits.
 We shall present a counterexample to this claim.

Let us take a state in the Pauli basis
$
|\vec{a}\rangle=((1+a_z)|0\rangle+(a_x+ia_y)|1\rangle)/2
$
so that the probability of the $p(a|c)=|\langle \vec{c}|\vec{a}\rangle|^2=(1+\vec{a}\cdot\vec{c})/2$.
Generalizing it to two independent qubits, we have  $p(a,b|c,d)=(1+\vec{a}\cdot\vec{c})(1+\vec{b}\cdot\vec{d})/4$.
We take the states and measurements as follows
\begin{align}
&\vec{c}_{0,1,2}=(1,0,0),\vec{c}_{3,4,5}=(0,1,0),\vec{c}_{6}=(0,0,1),\nonumber\\
&\vec{d}_{0,3,6}=(1,0,0),\:\vec{d}_{1,4}=(0,1,0),\:\vec{d}_{2,5}=(0,0,1),\nonumber\\
&\vec{a}_{0,2,4}=(1,0,0)=-\vec{a}_{1,3,5}=(-1,0,0),\:\nonumber\\
&\vec{a}_{6,8,10}=(0,1,0)=-\vec{a}_{7,9,11},\nonumber\\
&\vec{a}_{8,10,12}=(0,0,1)=-\vec{a}_{9,11,13},\nonumber\\
&\vec{b}_{0,6,12}=(1,0,0),\:\vec{b}_{2,8}=(0,1,0),\:\vec{b}_{4,10}=(0,0,1),
\end{align}
with $\vec{b}_{1,3,5}=-\vec{b}_{0,2,4}$ and $\vec{b}_{7,9,11,13}=\vec{b}_{6,8,10,12}$
Then 
\begin{align}
&W(i,j)=p(i,2j)-p(i,2j+1)=\nonumber\\
&\left\{\begin{array}{ll}
(\vec{c}_i\cdot\vec{a}_{2j}+\vec{d}_i\cdot\vec{b}_{2j})/2&\mbox{ for }j<3,\\
\vec{c}_i\cdot\vec{a}_{2j}(1+\vec{d}_i\cdot\vec{b}_{2j})/2&\mbox{ for }j\geq 3.
\end{array}
\right.
\end{align}
The matrix $p$ reads
\be
\begin{pmatrix}
1&1/2&1/2&0&0&0&0\\
1/2&1&1/2&0&0&0&0\\
1/2&1/2&1&0&0&0&0\\
1/2&0&0&1&1/2&1/2&0\\
0&1/2&0&1/2&1&1/2&0\\
0&0&1/2&1/2&1/2&1&0\\
1/2&0&0&1/2&0&0&1/2
\end{pmatrix},
\ee
whose determinant is $1/8$. It means that $W_7\neq 0$ contrary to the general claim of \cite{opt}.
We suspect that equality $W_7=0$ requires additional assumptions on preparations and measurements.

\begin{figure*}
\includegraphics[scale=1]{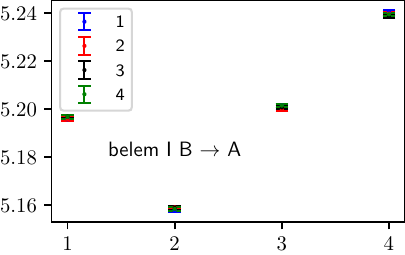}
\includegraphics[scale=1]{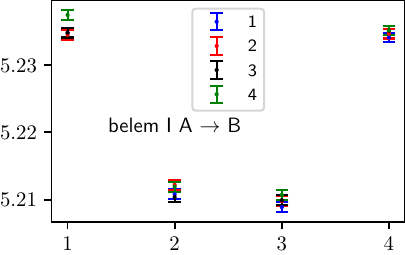}
\caption{No-signaling test for belem I, for (top) $p_{i0}$ ($i$ on the horizontal axis) depending on $j$ (inset),
and (bottom) $p_{0j}$ depending on $i$.}
\label{sig1belem}
\end{figure*}
\begin{figure*}
\includegraphics[scale=1]{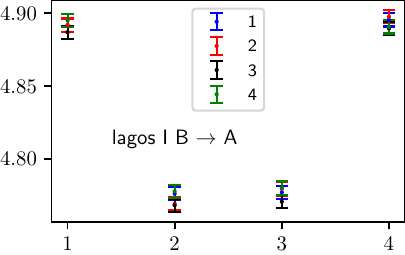}
\includegraphics[scale=1]{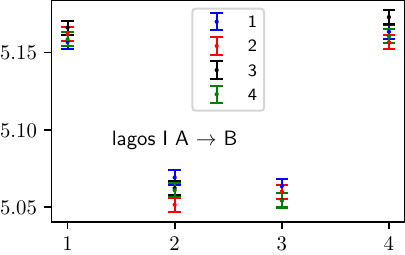}
\caption{No-signaling test for lagos I, notation as in Fig. \ref{sig1belem}}
\label{sig1lagos}
\end{figure*}
\begin{figure*}
\includegraphics[scale=1]{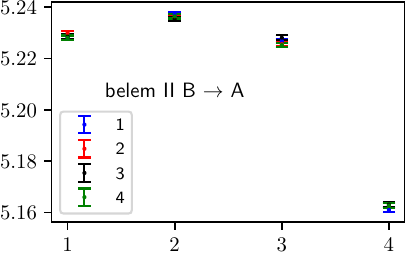}
\includegraphics[scale=1]{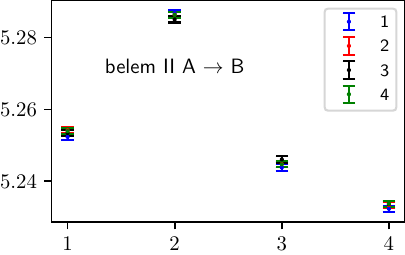}
\caption{No-signaling test for belem II, notation as in Fig. \ref{sig1belem}}
\label{sig2belem}
\end{figure*}
\begin{figure*}
\includegraphics[scale=1]{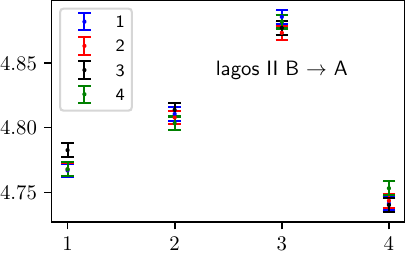}
\includegraphics[scale=1]{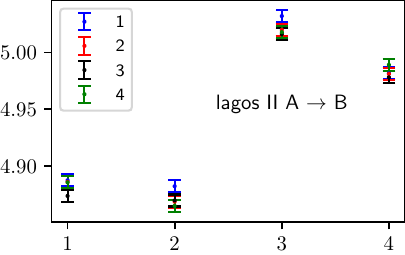}
\caption{No-signaling test for lagos II, notation as in Fig. \ref{sig1belem}}
\label{sig2lagos}
\end{figure*}
\begin{figure*}
\includegraphics[scale=1]{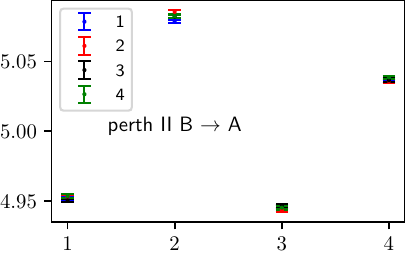}
\includegraphics[scale=1]{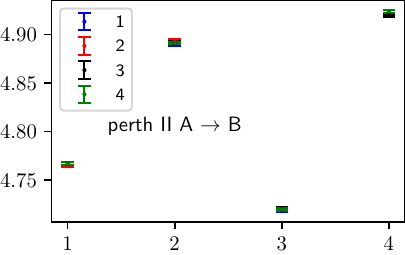}
\caption{No-signaling test for perth II, notation as in Fig. \ref{sig1belem}}
\label{sig2perth}
\end{figure*}
\begin{figure*}
\includegraphics[scale=1]{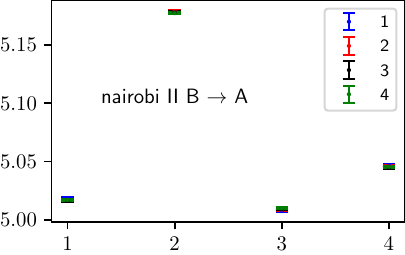}
\includegraphics[scale=1]{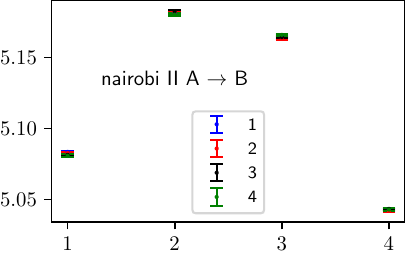}
\caption{No-signaling test for nairobi II, notation as in Fig. \ref{sig1belem}}
\label{sig2nairobi}
\end{figure*}

\section{No-signaling}
\label{appd}

The crucial assumption of the test (a) is locality. One can only verify a consequence of locality, namely no-signaling.
It means that the probability $p_{i0}$ and $p_{0j}$ can be found in any $(i,j)$ pair of measurements as $p_{i0,j}$ or $p_{0j,i}$, but does not depend on the other party,
i.e.
\be
p_{i0,j}=p_{i0},\:p_{0j,i}=p_{0j}.
\ee
In our demonstration, we could determine each $p_{i0,j}$ and $p_{0j,i}$, i.e. $2\times 4\times 4$ numbers.
The results are presented in Figs. \ref{sig1belem},\ref{sig1lagos},\ref{sig2belem},\ref{sig2lagos},\ref{sig2perth},\ref{sig2nairobi}.
In the case of belem, lagos, and nairobi, no-signaling is satisfied within the statistical error. However, for perth,  $p_{20,1}$ differs from $p_{20,2}$
by $3.9$ standard deviations, corresponding to $p$-value $10^{-4}$. The actual significance is lower because of 48 possible comparisons,
due to the look-elsewhere effect. Nevertheless, we suggest caution and repeating this test with different resources.

\section{Relation between $CNOT$ and $ECR$ gates}
\label{appe}

\begin{figure}
\begin{tikzpicture}[scale=1]
		\begin{yquantgroup}
			% q[1];
			\registers{
			qubit {} q[2];
			}
			\circuit{
			init {$a$} q[0];
			init {$b$} q[1];
			% cbit c[1];
			box {$\downarrow$}  (q[0,1]);}
			\equals
			\circuit{
			box {\rotatebox{90}{$CR^+$}}  (q[0,1]);
			box {$X$} q[0];
			box {\rotatebox{90}{$CR^-$}}   (q[0,1]);}
		\end{yquantgroup}
\end{tikzpicture}

\caption{The notation of the ECR gate in the convention $ECR_\downarrow|ab\rangle$}
\label{ecr}
\end{figure}
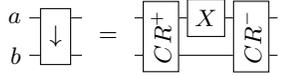
The IBM Quantum devices (brisbane) use a native two-qubit gate Echoed Cross Resonance ($ECR$) instead of $CNOT$.
However, one can transpile the latter by the former, adding single qubits gates.
We shall use Pauli matrices in the basis $|0\rangle$, $|1\rangle$,
\be
X=\begin{pmatrix}
0&1\\
1&0\end{pmatrix},\:Y=\begin{pmatrix}
0&-i\\
i&0\end{pmatrix},\:Z=\begin{pmatrix}
1&0\\
0&-1\end{pmatrix},\:
I=\begin{pmatrix}
1&0\\
0&1\end{pmatrix}.\label{pauli}
\ee
We also denote two-qubits gates by $\downarrow$ and $\uparrow$, which mean the direction of the gate (it is not symmetric), i.e.
$\langle a'b'|G_\uparrow|ab\rangle=\langle b'a'|G_\downarrow|ba\rangle$.

The $ECR$ gate acts on the states $|ab\rangle$ as (Fig. \ref{ecr})
\ba
&ECR_\downarrow=(XI-YX)/\sqrt{2}=CR^- (XI) CR^+=\nonumber\\
&
\begin{pmatrix}
0&X_-\\
X_+&0\end{pmatrix}
=\begin{pmatrix}
0&0&1&i\\
0&0&i&1\\
1&-i&0&0\\
-i&1&0&0\end{pmatrix}/\sqrt{2},
\ea
in the basis $|00\rangle$, $|01\rangle$, $|10\rangle$, $|11\rangle$
where the native gate is
\be
S=X_+=X_{\pi/2}=(I-iX)/\sqrt{2}=\begin{pmatrix}
1&-i\\
-i&1\end{pmatrix}/\sqrt{2},
\ee
and $X_-=X_{-\pi/2}=ZX_+Z$, 
with 
\be
CR^\pm=(ZX)_{\pm \pi/4},
\ee
using the convention  $V_\theta=\exp(-i\theta V/2)=\cos(\theta/2)-iV\sin(\theta/2)$ if $V^2=I$ or $II$.
The gate is its inverse, i.e. $ECR_\downarrow ECR_\downarrow=II$.

\begin{figure}
\begin{tikzpicture}[scale=1]
		\begin{yquantgroup}
			% q[1];
			\registers{
			qubit {} q[2];
			}
			\circuit{
			box {$\uparrow$} (q[0,1]);
			}
			\equals
			\circuit{
			% cbit c[1];
			box {$Y_+$}  q[0];
			box {$Y_-$}   q[1];
			box {$\downarrow$}  (q[0,1]);
			box {$H$}  q[0];
			box {$H$}   q[1];
			}
		\end{yquantgroup}
\end{tikzpicture}

\caption{The $ECR_\uparrow$ gate expressed by $ECR_\downarrow$ }
\label{ecrr}
\end{figure}
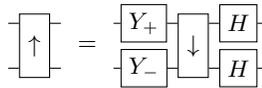

Note that $Z_\theta=\exp(-i\theta Z/2)=\mathrm{diag}(e^{-i\theta/2},e^{i\theta/2})$ is a virtual gate adding essentially the phase  shift to next gates.
 $ECR$ gates can be reversed, i.e., for $a\leftrightarrow b$, (Fig. \ref{ecrr})
\be
ECR_\uparrow=(IX-XY)/\sqrt{2}=(HH) ECR_\downarrow(Y_+Y_-),
\ee
denoting $V_\pm= V_{\pm \pi/2}$, and Hadamard gate,
\be
H=(Z+X)/\sqrt{2}
=Z_+SZ_+=\begin{pmatrix}
1&1\\
1&-1\end{pmatrix}/\sqrt{2},
\ee
and $Z_\pm SZ_\mp=Y_\pm$, with $Y_+=HZ$ and $Y_-=ZH$.

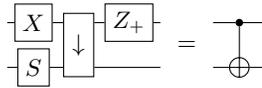
\begin{figure}
\begin{tikzpicture}[scale=1]
		\begin{yquantgroup}
			% q[1];
			\registers{
			qubit {} q[2];
			}
			\circuit{
			% cbit c[1];
			box {$X$}  q[0];
			box {$S$}   q[1];
			box {$\downarrow$}  (q[0,1]);
			box {$Z_+$}  q[0];
			}
			\equals
			\circuit{
			cnot q[1] | q[0];
			}
		\end{yquantgroup}
\end{tikzpicture}

\caption{The $CNOT_\downarrow$ gate expressed by $ECR_\downarrow$ }
\label{cnot}
\end{figure}

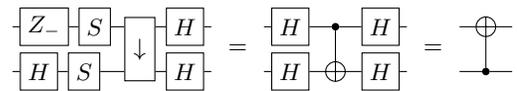
\begin{figure}
\begin{tikzpicture}[scale=1]
		\begin{yquantgroup}
			% q[1];
			\registers{
			qubit {} q[2];
			}
			\circuit{
			% cbit c[1];
			box {$Z_-$}  q[0];
			box {$H$}   q[1];
			box {$S$}  q[0];
			box {$S$}   q[1];
			box {$\downarrow$}  (q[0,1]);
			box {$H$}  q[0];
			box {$H$}  q[1];
			}
			\equals
			\circuit{
			box {$H$}  q[0];
			box {$H$}   q[1];
			cnot q[1] | q[0];
			box {$H$}  q[0];
			box {$H$}   q[1];
			}
			\equals
			\circuit{
			cnot q[0] | q[1];
			}
		\end{yquantgroup}
\end{tikzpicture}

\caption{The $CNOT_\uparrow$ gate expressed by $ECR_\downarrow$ }
\label{cnotr}
\end{figure}

The $CNOT$ gate can be expressed by $ECR$ (Fig. \ref{cnot})
\ba
&CNOT_\downarrow=(II+ZI+IX-ZX)/2=\nonumber\\
&
\begin{pmatrix}
I&0\\
0&X\end{pmatrix}
=\begin{pmatrix}
1&0&0&0\\
0&1&0&0\\
0&0&0&1\\
0&0&1&0\end{pmatrix}\nonumber\\
&
=(Z_+ I)ECR_\downarrow (XS),
\ea
while its reverse reads (Fig. \ref{cnotr})
\ba
&CNOT_\uparrow=(II+IZ+XI-XZ)/2
=\nonumber\\
&\begin{pmatrix}
1&0&0&0\\
0&0&0&1\\
0&0&1&0\\
0&1&0&0\end{pmatrix}=(HH)CNOT_\downarrow(HH)\nonumber\\
&
=
(HH)ECR_\downarrow (SS)(Z_-H).
\ea

\section{Experimental data}
\label{appf}

\begin{figure*}
\includegraphics[scale=1]{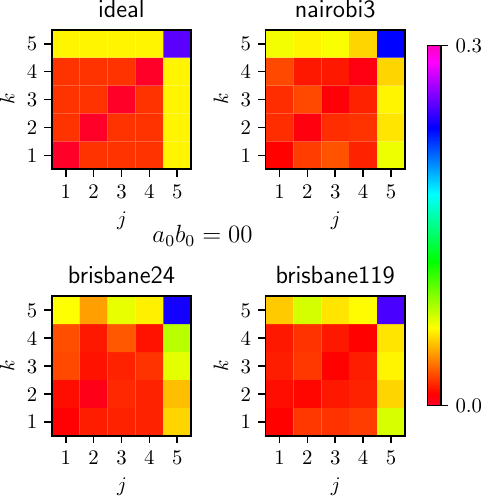}
\includegraphics[scale=1]{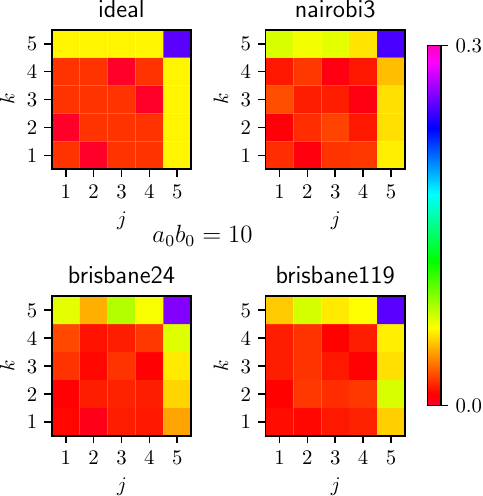}
\includegraphics[scale=1]{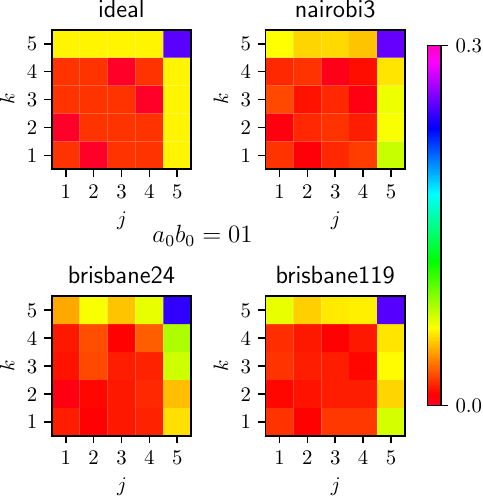}
\includegraphics[scale=1]{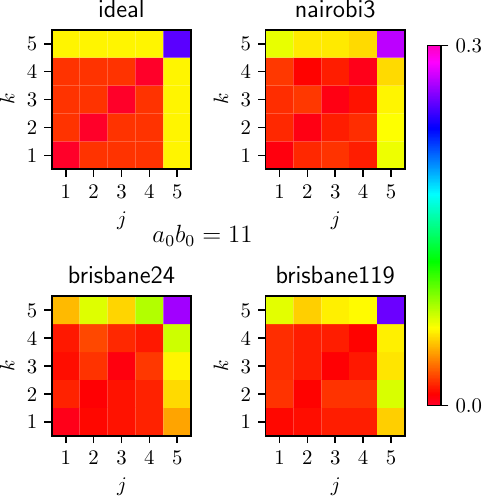}
\caption{The probabilities for the test (b) for each combination of spectator qubits $a_0b_0$, see main text, with the middle qubit 3 (nairobi), 24 and 119 (brisbane), compared to the ideal theoretical expectation.}
\label{meshx}
\end{figure*}

\begin{figure*}
\includegraphics[scale=1]{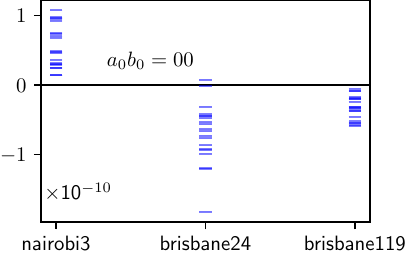}
\includegraphics[scale=1]{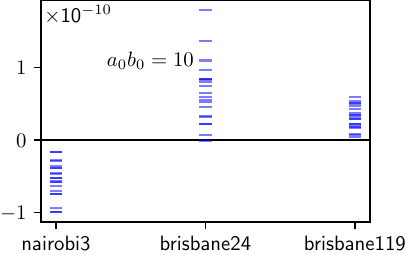}
\includegraphics[scale=1]{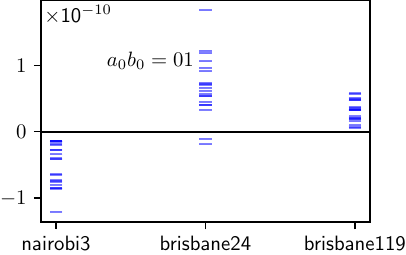}
\includegraphics[scale=1]{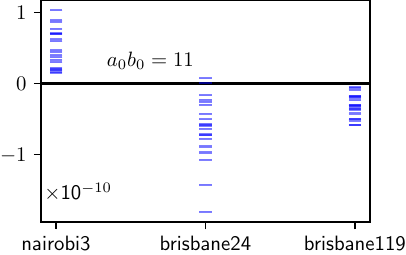}
\caption{The results of the witness calculated for individual jobs in the test (b) for each combination of chosen spectator qubits values $a_0b_0$, see main text, with the middle qubit 3 (nairobi), 24, and 119 (brisbane)}
\label{scb}
\end{figure*}

The technical characteristics of the relevant qubits and two-qubit connecting gates are shown in Table \ref{tech} and \ref{tech2}.
the actual maps of probabilities in the case of test (b) are shown in Fig. \ref{meshx}.
The results of the witness in the case (b) for individual jobs are presented in Fig. \ref{scb}.

\section*{Abbreviations}
ECR, echoed cross-resonance.
\section*{Declarations}
\subsection*{Ethical Approval and Consent to participate}
Not applicable.
\subsection*{Consent for publication}
Not applicable.
\subsection*{Availability of supporting data}
The data are publicly available at 
https://doi.org/10.5281/zenodo.8358855
\subsection*{Competing interests}
The authors declare no competing interests.
\subsection*{Funding}
TR gratefully acknowledges the funding support by the
program ,,Excellence initiative research university'' for the AGH University in
Krakow as well as the ARTIQ project: UMO-2021/01/2/ST6/00004 and
ARTIQ/0004/2021.
\subsection*{Authors' contributions}
J.B and A.B developed the idea. T.B. and T.R. wrote computer scripts. J.T. and A.B. wrote the manuscript.
All the Authors reviewed the manuscript.

\end{document}